\documentclass
[twocolumn,showpacs,10pt,notitlepage,tightenlines,lengthcheck,superscriptaddress,prl]{revtex4}%
\usepackage{amsfonts}
\usepackage{amsmath}
\usepackage{amssymb}
\usepackage{graphicx}%
\providecommand{\U}[1]{\protect\rule{.1in}{.1in}}
\begin{document}
\title{A Single Charged Quantum Dot in a Strong Optical Field: Absorption, Gain, and
the AC Stark Effect}
\author{Xiaodong Xu}
\author{Bo Sun}
\author{Erik D. Kim}
\author{Katherine Smirl}
\author{P. R. Berman}
\author{D. G. Steel}
\email{dst@umich.edu}
\affiliation{The H. M. Randall Laboratory of Physics, The University of Michigan, Ann
Arbor, Michigan 48109, USA}
\author{A. S. Bracker}
\author{D. Gammon}
\affiliation{The Naval Research Laboratory, Washington D.C. 20375, USA}
\author{L. J. Sham}
\affiliation{Department of Physics, The University of California-San Diego, La Jolla,
California 92093, USA}
\date{\today}

\begin{abstract}
We investigate a singly-charged quantum dot under a strong optical
driving field by probing the system with a weak optical field. When
the driving field is detuned from the trion transition, the probe
absorption spectrum is shifted from the trion resonance as a
consequence of the dynamic Stark effect. Simultaneously, a gain
sideband is created, resulting from the coherent energy transfer
between the optical fields through the quantum dot nonlinearity. As
the pump detuning is moved from red to blue, we map out the
anticrossing of these two spectral lines. The optical Bloch
equations for a stationary two-level atom can be used to describe
the numerous spectral features seen in this nano solid state system.

\end{abstract}

 \maketitle

Quantum dot (QD) nano-structures have been proposed for numerous quantum
mechanical applications due to their customizable atom-like features
\cite{DanPhysToday}. One important application involves using these QDs as the
building blocks for quantum logic devices \cite{D Loss' PRA}. An electron spin
trapped inside a QD is a good candidate for a quantum bit (qubit) since it is
known to have long relaxation \cite{Miro's nature} and decoherence times
\cite{Macus's spin echo, Bayer's mode locking}. Recently, the electron spin
coherence has been optically generated and controlled \cite{Bayer's mode
locking, Dutt's PRL,Bayer's Nuclear Spin focusing} in ensembles of QDs. The
initialization of the electron spin state in a single QD has also been
realized by optical cooling techniques \cite{Imamoglu's optical pumping,
XD'sOpticalPumping}.

One important task is to understand and control the physical
properties of a singly-charged QD in the strong optical field
regime, i.e. the light-matter interaction strength is much larger
than the transition linewidth, under both resonant and nonresonant
excitation. Given the recent work on optically driven neutral
quantum dots in strong fields \cite{XDMollow,ShihTriplets}
demonstrating many features similar to atomic systems, it is clear
that a negatively charged quantum dot has similarities to a negative
ion.  However, the excited state of a dot is a many body structure
comprised of two electrons and a hole. Interestingly, the results in
this paper show that strong field excitation tuned near resonance in
a negatively charged dot leads to changes in the absorption spectrum
that are in excellent agreement with theory.

In the time domain, the strong field interaction leads to the
well-known Rabi oscillations
\cite{Tod2001PRL,Kamada2001PRL,Shih2002PRL,ZrennerNature}. In the
frequency domain, it will introduce Rabi side bands in the
absorption, and strikingly, the amplification of a probe beam. This
phenomenon has been studied theoretically
\cite{Baklanov,MollowSpectrum,Haroche} and demonstrated
experimentally in atomic systems \cite{Ezekiel,Boyd}. Recently,
these effects have also been observed in quantum dot and molecular
systems \cite{ACStark,XDMollow,ShihTriplets,Wrigge}. The optical AC
Stark effect has been seen by exciting a neutral QD with a detuned
strong optical pulse \cite{ACStark} while the Mollow absorption
spectrum \ and Mollow triplets \cite{Mollowtriplets} have been
observed in a single neutral QD \cite{XDMollow, ShihTriplets} and a
single molecule with intense resonant pumping \cite{Wrigge},
respectively. However, the study of the singly-charged QDs in the
strong field regime has been very limited at the single dot
level~\cite{WAPL}.

\begin{figure}[ptb]
\centerline{ \scalebox{.50} {\includegraphics{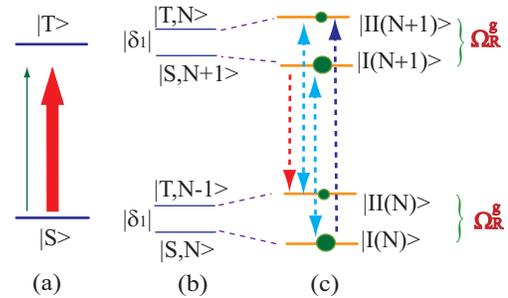}}}
\caption{(color online) (a) The energy level diagram of a trion
state at zero magnetic field. The absorption spectrum of the weak
probe (green arrow) is modified by a strong pump field (red arrow).
(b) The uncoupled atom-field states. (c) Dressed state picture of a
two-level system driven by a strong optical field. The energy levels
outside the picture are not shown. The energy splitting
between the dressed states with the same photon number is $\hbar\Omega_{R}%
^{g}$, where $\Omega_{R}^{g}$ is the generalized Rabi frequency.}%
\label{fig. 5}%
\end{figure}

In this letter, we investigate a singly-charged QD under a strong
optical driving field with both on and off-resonant pumping. When
the strong pump is on resonance with the trion transition, a triplet
appears in the probe absorption spectrum with a weak center peak and
two Rabi side bands with dispersive lineshapes. As the pump beam is
detuned from the trion transition, we observe three spectral
features: a weak dispersive lineshape centered at the driving field
frequency flanked by an AC Stark shifted absorption peak and a Raman
gain side band. Our results reflect the coherent nonlinear
interaction between the light and a single quantum oscillator, and
demonstrate that even at high optical field strengths, the electron
in a single quantum dot with its ground state and trion state
behaves as a well isolated two-level quantum system. It is a step
forward toward spin based QD applications.

Assuming the trion can be considered as a simple two-level system in
the absence of the magnetic field, the only optically allowed
transitions are from the spin ground states ($\left\vert
\pm\frac{1}{2}\right\rangle $) to the trion states ($\left\vert
\pm\frac{3}{2}\right\rangle $) with $\sigma\pm$ polarized light
excitations. Since the Zeeman sub-levels of the electron spin ground
state are degenerate, as are the trion states, both trion
transitions are degenerate. We then use the two-level optical Bloch
equations to model the trion system. For simplicity, we labeled the
electron spin ground state as state $\left\vert S\right\rangle $ and
the excited state as $\left\vert T\right\rangle $, as shown in Fig 1
(a).

It is known that in a two-level system driven by a strong optical
field, the absorption of the weak probe beam is significantly
modified \cite{Baklanov,MollowSpectrum,Haroche}. By solving the
optical Bloch equations to all orders in the pump field and first
order in the probe field, we obtain the absorption coefficient of
the probe beam as \cite{MollowSpectrum}

\begin{widetext}
\begin{equation}
\alpha=\operatorname{Im}[\frac{%
\begin{array}
[c]{c}%
\alpha_{o}\gamma_{T}A[i\gamma_{T}A^{\ast}(A^{\ast}+i\Delta)+
\Delta(\frac{\Omega_{R}^{2}}{2}-\gamma B+i\left(  B+i\Delta\right)
\delta _{1}+\delta_{1}^{2})]
\end{array}
}{%
\begin{array}
[c]{c}%
(\gamma_{T}^{2}AA^{\ast}+i\Omega_{R}^{2}\gamma\Delta)(B^{2}+\delta_{1}^{2})+
\gamma\Omega_{R}^{4}B+\gamma_{T}(\Omega_{R}^{2}(\gamma+B)(\gamma
B+\delta _{1}^{2})+
i\gamma^{2}B^{2}\Delta+i\delta_{1}^{2}\Delta(B^{2}+\gamma^{2}+\delta_{1}^{2}))
\end{array}
}], \label{probeabsorption}%
\end{equation}
\end{widetext}
where $\gamma_{T}$ ($\gamma$) is the population (coherence) decay
rate of the trion state, $\delta_{1}=\omega_{1}-\omega_{o}$ is the
detuning of the pump frequency ($\omega_{1}$) from the trion
transition ($\omega_{o}$), $\Delta=\omega_{2}-\omega_{1}$ is the
probe ($\omega_{2}$) detuning from the pump, $A=\gamma+i\delta_{1}$,
$B=\gamma+i\Delta,\alpha_{o}$ is a constant, $\Omega_{R}={\mu\cdot
E_\text{pump}}/{\hbar}$ is the Rabi frequency of the pump field,
$\mu$ is the dipole moment matrix element and $E_\text{pump}$ is the
pump field strength.

When the strong pump is on resonance with the trion transition ($\delta_{1}=0$
and $\Omega_{R}>>\gamma$), the probe will show a complex Mollow absorption
spectrum, which has been discussed in detailed in Ref \cite{XDMollow}, where a
neutral exciton has been studied with a strong resonant pumping.

When the pump detuning is larger than the transition linewidth, the
physics can be understood in the fully quantized dressed state
picture. The uncoupled QD-field states (Fig. \ref{fig. 5}(b)) map
into the dressed states (Fig.  \ref{fig. 5}(c)) when the QD-field
interaction is included. In Fig.  \ref{fig. 5}, we assume the pump
detuning $\delta_{1}$ to be negative, $\left\vert S\right\rangle $
and $\left\vert T\right\rangle $ are the quantum dot states, and $N$
is the photon number. Due to the light-matter interaction, one set
of the dressed states can be written as \cite{CTBook}
\begin{align*}
\left\vert I(N)\right\rangle  &  =c\left\vert S,N\right\rangle -s\left\vert
T,N-1\right\rangle \\
\left\vert II(N)\right\rangle  &  =s\left\vert S,N\right\rangle +c\left\vert
T,N-1\right\rangle
\end{align*}
where $c=\sqrt{\frac{1}{2}\left(  1-\frac{\delta_{1}}{\Omega_{R}^{g}}\right)
}$, $s=\sqrt{\frac{1}{2}\left(  1+\frac{\delta_{1}}{\Omega_{R}^{g}}\right)  }%
$, and $\Omega_{R}^{g}=\sqrt{\Omega_{R}^{2}+\delta_{1}^{2}}$ is the
generalized Rabi frequency. The energy separation between the
dressed states $\left\vert I(N)\right\rangle $ and $\left\vert
II(N)\right\rangle $ is $\hbar\Omega_{R}^{g}$. As shown in Fig.
1(c), there are three transition frequencies: one centered at the
pump frequency $\omega_{1}$, and two Rabi side bands centered at
frequency $\omega_{1}\pm\Omega_{R}^{g}$.

Assuming $\Omega_{R}^{g}>>\gamma$ and using the secular
approximation, the steady state solutions for the dressed state
population are
\[
\rho_{I,I}=\frac{c^{4}}{c^{4}+s^{4}},\rho_{II,II}=\frac{s^{4}}{c^{4}+s^{4}}.
\]
It is clear when $\delta_{1}$ $<0$, the dressed state $\left\vert
I(N)\right\rangle $ is more populated than the dressed state
$\left\vert II(N)\right\rangle $. In Fig. 1(c), the size of the dots
on states $\left\vert I(N)\right\rangle $ $(\left\vert
I(N+1)\right\rangle $ and $\left\vert II(N)\right\rangle $
$(\left\vert II(N+1)\right\rangle $ indicates their populaiton.
Therefore, the transition centered at $\omega_{1}+\Omega_{R}^{g}$
represents probe absorption (the purple dashed line in Fig. 1(c)),
and the transition centered at $\omega_{1}-\Omega_{R}^{g}$ is probe
gain due to the population inversion of the dressed states (the red
dashed line in Fig. 1(c)). The gain process, in its simplest form,
can also be considered as a three photon process, in which two pump
photons are absorbed at frequency $\omega_{1}$ and a third photon is
emitted at frequency $\omega_{1}-\Omega_{R}^{g}$ \cite{Boyd}. The
light blue lines indicate transitions where the probe frequency is
close to the pump frequency and the secular approximation fails.
These can give rise to a dispersive lineshape \cite{Boyd,central
resonance}.

The experiment is performed on a singly-charged self-assembled InAs
QD embedded in a Schottky diode structure. The detailed sample
information can be found in Ref \cite{Morgan's PRL,XDMollow}. The
sample is located in a continuous helium flow magneto cryostat at a
temperature of $5$ K. By varying the DC gate voltage across the
sample, the charge state of the dot can be controlled
\cite{warburton,Morgan's PRL} and the transition energies can be
electrically tuned using the DC Stark effect \cite{Karrai's APL}.
When the DC Stark shift is modulated by a small AC voltage, the
changes in the transmission signal can be detected at the modulation
frequency by a phase-sensitive lock-in amplifier.

By setting the voltage modulation amplitude to about $16$ times the transition
linewidth, we avoid complexities associated with smaller modulations
\cite{Karrai's APL}. The data taken directly correspond to the absorption. To
obtain the Mollow absorption spectrum, two continuous wave (CW) lasers are
used. In the pump-probe experiment, we set both beams to be linearly polarized
with orthogonal polarization. By filtering out the pump beam with a polarizer
in front of the detector, we can measure the probe absorption only.

We first set the pump detuning $\delta_{1}$ to be zero and scan the
probe frequency across the trion transition frequency $\omega_{o}$.
Figure 2(a) shows the probe absorption lineshape with a pump
intensity of $95$ $W/cm^{2}$. Instead of a lorentzian absorption
lineshape in the absence of the pump, as shown at the bottom of the
Fig. 2(a), the lineshape of the probe beam in the presence of a
strong pump beam shows a complex structure \cite{note}: a
triplet-like absorption pattern appears with one weak central
structure and two Rabi side bands with dispersive lineshape. The
observation of the Rabi side bands is a signature of the optical
generation of a single dot trion Rabi oscillations. The inset in
Fig. 2(a) shows the Rabi spitting of the side bands as a function of
the pump intensity. The largest Rabi frequency we achieved in the
experiment is about 2$\pi\times$1.6 GHz, limited only by the current
experimental configuration.

\begin{figure}[ptb]
\centerline{ \scalebox{1.1} {\includegraphics{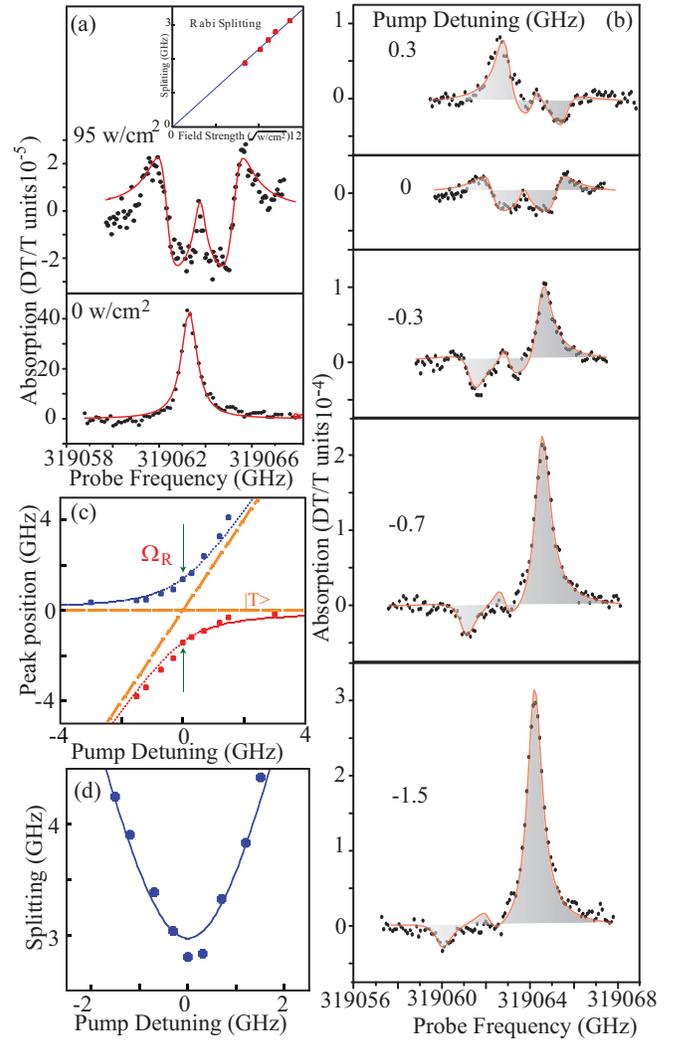}}}
\caption{(color online)(a) Top curve: trion Mollow absorption
spectrum at a pump intensity of $95$ $W/cm^{2}$ with resonant
pumping. Bottom curve: a probe beam absorption spectrum with no
pump. Inset: the Rabi splitting of the side bands as a function of
the pump intensity. (b) Trion Mollow absorption spectrum with
various pump detuning with a fixed pump intensity of $95$
$W/cm^{2}$. Two Rabi side bands are clearly observed, where one is
the AC Stark shifted absorption peak and the other shows gain. (c)
The spectral position of the Rabi side bands as a function of the
pump detuning. We use the trion transition energy as the zero point.
The anti crossing feature of the Rabi side bands is demonstrated as
the pump is detuned from the red to the blue of the trion
transition. (d) The energy separation of the Rabi side bands as a
function of the pump detuning. The solid blue line is the fits by
the formula $2\sqrt{\delta_{1}^{2}+\Omega
_{R}^{2}}$.}%
\label{fig. 2}%
\end{figure}

The negative part of the absorption lineshape demonstrates gain of
the probe beam. Since the pump is resonant with the trion
transition, there is no population inversion in the steady state of
the trion system in any picture. The gain effect comes from the
coherent energy exchange between the pump and probe beams through
the QD nonlinearity. We define the efficiency of the probe gain as
the ratio of the amplitude of the negative absorption to the probe
absorption in the absence of the strong pump. The probe gain
efficiency corresponding to a pump intensity of $95$ $W/cm^{2}$ is
$5.3$ $\%$.  The earlier work by Kroner \textit{et. al.}~\cite{WAPL}
in a negatively charged dot did not show gain in their spectrum
though they saw many of the other features consistent with strong
field excitation.  They attribute this difference to possible
effects of dephasing.

As we tune the pump laser frequency away from the trion transition, the
dispersion-like lineshapes of the Rabi side bands evolve into three spectral
features: one weak central structure with a dispersive lineshape and two Rabi
side bands with Lorentzian lineshapes. Figure 2(b) displays the probe
absorption spectrum as a function of the pump detuning with a fixed pump
intensity of $95$ $W/cm^{2}$.

A distinct feature of the probe absorption spectrum is that one of
the side bands shows purely negative ``absorption", which is the
gain effect. Using the pump detuning at $-1.5$ $GHz$ as an example
(the bottom curve of Fig.
\ref{fig. 2}(b)), there is an absorption peak located at $\omega_{1}%
+\Omega_{R}^{g}$ $.$ This is an AC stark shifted absorption peak.
The side band centered at $\omega_{1}-\Omega_{R}^{g}$ is negative,
which signifies the amplification of the probe beam. In lowest order
perturbation theory, this reflects a three photon Raman gain effect:
the QD absorbs two pump photons at frequency $\omega_{1}$ and emits
a photon at $\omega_{1}-\Omega_{R}^{g}$ . The frequency at which
gain occurs can be tuned by adjusting the pump detuning. As
expected, if the pump detuning is positive, the probe sees gain at
$\omega_{1}+\Omega_{R}^{g}$. The data with pump detuned $+0.3$ $GHz$
is shown at the top of Fig. \ref{fig. 2}(b). A gain peak is clearly
observed for the positive detuning of the probe. It has been shown
theoretically that the maximum gain occurs at the absolute value of
the pump detuning $\left\vert
\delta_{1}\right\vert =\Omega_{R}^{g}/3$ provided $\Omega_{R}%
^{g}>>\gamma$ \cite{theoryGainEffi}. For the pump detuning $-0.3$ $GHz$, the
data shows a probe gain of $9.7$ $\%$, which is much larger than under
resonant pumping with the same intensity. When the probe frequency is nearly
degenerate with the pump beam, there is also a small dispersive structure in
the probe absorption spectrum, as shown in Fig. 2(b).

The solid lines in Fig. \ref{fig. 2}(b) are theoretical fits of the
data to Eq.~(1). The fits yield $\gamma_{T}/2\pi$ and $\gamma/2\pi$
of $(580\pm90)$ $\text{MHz}$ and $(350\pm35)$ $\text{MHz}$,
respectively. Since $\gamma_{T}$ is almost twice $\gamma$, the
amount of pure dephasing in this QD is statistically insignificant
compared with the error bars. These fits show that our results can
be well reproduced by the Optical Bloch equations and that the
singly charged QD behaves like a single isolated atomic system.

Figure \ref{fig. 2}(c) shows the spectral positions of the Rabi side
bands as a function of the pump detuning. In the plot, we use the
trion transition frequency $\omega_{o}$ as the zero energy point.
Figure \ref{fig. 2}(c) clearly illustrates the anti crossing
behavior of the Rabi side bands. The separation between the two
peaks at zero pump detuning represents the interaction strength
between the light and QD, equal to the Rabi frequency. The dotted
curves in the plot are the theoretical predictions of the peak
positions as a function of the detuning, which is in good agreement
with the measurements. The laser light induced transition energy
shifts at the large pump detuning are a demonstration of the
dynamic, or AC stark effect.

We extracted the energy separation of the side bands from the data
and plotted it as a function of the pump detuning in Fig. \ref{fig.
2}(d). The solid blue line is the fit by the expression
$2\sqrt{\Omega_{R}^{2}+\delta_{1}^{2}}$ and gives
$\Omega_{R}/2\pi=(1.5\pm0.1)$ $GHz$. \ Since $\Omega_{R}=\mu\cdot
E_\text{pump}/{\hbar}$, we infer the trion dipole moment of
$(25\pm2)$ $D$. The trion dipole moment\ we calculated is similar to
the reported neutral exciton dipole moment \cite{XDMollow}.

The Einstein A coefficient, or spontaneous emission rate, is
\cite{EinsetinACoef}
\begin{align}
\gamma_\text{sp} &  =\frac{9n^{5}}{(2n^{2}+n_\text{QD}^{2})^{2}}\frac{\omega_{o}^{3}%
\mu^{2}}{3\pi\varepsilon_{o}\hbar c^{3}}\label{einstein A coefficient}\\
&
=\frac{9n^{5}}{(2n^{2}+n_\text{QD}^{2})^{2}}\gamma_\text{spo}\nonumber
\end{align}
where $\gamma_\text{spo}$ is the spontaneous emission rate in the
vacuum, $n$ and $n_\text{QD}$ are the refractive index of the medium
and the QD, respectively. By inserting the parameters into
Eq.~(\ref{einstein A coefficient}), \ we get a spontaneous emission
rate of $2\pi\times130$ $\text{MHz}$, which corresponds to a trion
radiative life time of 1.2 $\text{ns}$. Assuming there is no pure
dephasing in the QD, as we shown earlier, then the trion transition
linewidth is about 130 $\text{MHz}$, which is smaller than what we
extracted from our previous fits. Also, the low power single beam
absorption data yields a transition linewidth of $600$ $\text{MHz}$,
which is much larger than what we calculated from the Einstein A
coefficient. This discrepancy could come from the spectral diffusion
process, which broadens the trion transition
linewidth\cite{XDMollow, Hogele's PRL}.

In summary, we have shown that an electron trapped inside a QD with
its ground state and the excited two electron and one hole state
behaves as an isolated quantum system even in the strong field limit
by observing the optical Mollow absorption spectrum as well as the
AC Stark effect. The behavior is well described by the solutions to
the optical Bloch equations for a two-level system and show that the
state of the electron can be switched at a rate of 2$\pi\times$1.6
GHz with low power cw diode lasers.

This work is supported by the U.S. ARO, AFOSR, ONR, NSA/LPS, and
FOCUS-NSF.


\begin{thebibliography}{99}
\bibitem {note}There is a small energy shift of the weak probe absorption
spectra between the presence and the absence of the pump, which is probably
due to a small screening of the applied field by photoexcited charge in the
diode. This shift saturates at a power between the lowest-intensity curve and
the next higher-power spectrum. In the experiment, the pump laser is adjusted
to follow the shift of the resonance.
\end{thebibliography}
\end{document}